\documentclass[12pt]{article}

\usepackage{tikz}
\usetikzlibrary{external}
\usetikzlibrary{trees, mindmap, positioning}
\usepackage{amsfonts}
\usepackage{amsmath}
\usepackage{hyperref}
\usepackage{graphicx}
\usepackage{xcolor}
\usepackage{appendix}
\usepackage{lineno}

\date{}

\newcommand{\BR}{\mathcal{BR}}
\newcommand{\NBR}{\mathcal{NBR}}
\newcommand{\argmax}{\text{argmax}}

\begin{document}

\title{Evolution favours positively biased reasoning in sequential interactions with high future gains}

\author{
	Marco Saponara$^{1,a,\ast}$,
	Elias Fernández Domingos$^{1,2,b}$,\and
	Jorge M. Pacheco$^{3,4,c}$,
    Tom Lenaerts$^{1,2,5,d}$\and
	\small$^{1}$Machine Learning Group, ULB, Campus la Plaine, Brussels 1050, Belgium\and
	\small$^{2}$Artificial Intelligence Lab, VUB, Pleinlaan 9, Brussels 1050, Belgium\and
    \small$^{3}${INESC-ID, IST-Tagusparque, 2744-016 Porto Salvo, Portugal}\and
    \small$^{4}${ATP-group, P-2744-016 Porto Salvo, Portugal}\and
    \small$^{5}$Center for Human-Compatible AI, UC Berkeley, 2121 Berkeley Way, Berkeley, CA 94720, USA\and
    \small$^{a}$ORCID iD: 0009-0008-8092-6220\and
    \small$^{b}$ORCID iD: 0000-0002-4717-7984\and
    \small$^{c}$ORCID iD: 0000-0002-2579-8499\and
    \small$^{d}$ORCID iD: 0000-0003-3645-1455\and
	\small$^\ast$Corresponding author. Email: marco.saponara@ulb.be 
}



\maketitle

\begin{abstract}
Empirical evidence shows that human behaviour often deviates from game-theoretical rationality.
For instance, humans may hold unrealistic expectations about future outcomes.
As the evolutionary roots of such biases remain unclear, we investigate here how reasoning abilities and cognitive biases co-evolve using Evolutionary Game Theory.
In our model, individuals in a population deploy a variety of unbiased and biased level--$k$ reasoning strategies to anticipate others’ behaviour in sequential interactions, represented by the Incremental Centipede Game.
Positively biased reasoning strategies have a systematic inference bias towards higher but uncertain rewards, while negatively biased strategies reflect the opposite tendency.
We find that selection consistently favours positively biased reasoning, with rational behaviour even going extinct.
This bias co-evolves with bounded rationality, as the reasoning depth remains limited in the population.
Interestingly, positively biased agents may co-exist with non-reasoning agents, thus pointing to a novel equilibrium.
Longer games further promote positively biased reasoning, as they can lead to higher future rewards. 
The biased reasoning strategies proposed in this model may reflect cognitive phenomena like wishful thinking and defensive pessimism.
This work therefore supports the claim that certain cognitive biases, despite deviating from rational judgment, constitute an adaptive feature to better cope with social dilemmas.
\textbf{This is the early version of a manuscript published in the Royal Society Interface journal. DOI:} \url{https://doi.org/10.1098/rsif.2025.0153}
\end{abstract}




\section{Introduction}

A large body of experimental evidence has shown that humans can behave differently from the self-interested rational decision makers hypothesized by game-theoretical models~\cite{camerer_behavioral_2011, mckelvey_experimental_1992, rand_social_2014, oosterbeek_cultural_2004, grosskopf_two-person_2008}.
In fact, our decisions may not produce outcomes that align with classical solution concepts such as the Nash equilibrium \cite{camerer2004behavioural}. Instead, they may contain biases, i.e., consistent deviations from a reference -- rational behaviour -- deemed to be correct~\cite{jensen_chimpanzees_2007, bornstein_rationality_2001, montibeller_cognitive_2015}.
Such biases, on the one hand, influence strategic reasoning and, on the other hand, render their evolutionary roots difficult to explain~\cite{marshall_evolutionary_2013}.


Trivers' \textit{theory of self-deception}~\cite{trivers_deceit_2011, von_hippel_evolution_2011} argues that the evolution of cognitive biases stems from the adaptive advantages of deceiving oneself to better persuade or deceive others, thereby reducing the cognitive cost associated with dishonesty in social interactions.
This capacity is closely related to the notion of  \textit{Theory of Mind} (ToM)~\cite{rusch_theory_2020}, which allows us to attribute mental states such as beliefs, intentions, and desires to others~\cite{mckay_evolution_2009}. 
In fact, this early process enabled the development of a crucial feature of human social cognition that is usually defined as \textit{inter-subjectivity} (i.e., an understanding of the intentionality of others), thus rendering social transgressions more likely to be detected and punished.

In a recent publication~\cite{lenaerts_evolution_2024}, we proposed an evolutionary model to study the conditions under which ToM may have evolved, finding that intermediate levels of ToM, interpreted as recursive reasoning of the sort \textit{A thinks that B thinks that A thinks that B will do X}~\cite{camerer_cognitive_2004,rusch_theory_2020}, co-evolve with an optimism bias regarding the expected behaviour of other individuals in the context of mixed-motive interactions. 
This theoretical work also showed that the expectation of higher future rewards is essential for the appearance of ToM, and that the behaviour generated by the surviving reasoning strategies aligns well with previous experimental data~\cite{mckelvey_experimental_1992}.


Here, we go beyond the work in~\cite{lenaerts_evolution_2024} by exploring the evolutionary dynamics of biased and unbiased reasoning strategies using the framework of Evolutionary Game Theory~\cite{smith_logic_1973,smith_evolution_1982,sigmund_calculus_2010,fernandez_domingos_egttools_2023}.
Strategic reasoning is represented, as in \cite{lenaerts_evolution_2024}, through the lens of level--$k$ theory~\cite{stahl_evolution_1993,kawagoe_level-k_2012, nax_deep_2022}, a popular approach to model \textit{bounded rationality}~\cite{aumann_rationality_1997} (i.e., the idea that people have limited cognitive capacity and might follow a strategy which is not perfectly rational).
In this framework, the degree of rationality, or \textit{sophistication}, of an individual is represented by an integer $k\ge 0$, where no reasoning ($k=0$) corresponds to an individual acting on ingrained beliefs while higher values translate into a more sophisticated reasoning capacity.

The goal here is to investigate whether reasoning strategies with particular properties may be favoured by evolution.
In our evolutionary model, described in \autoref{sec:methods}, individuals are equipped with reasoning strategies whose
cognitive bias within the reasoning process is modulated by a noise parameter $\varepsilon$. Depending on the type of reasoning process, the parameter $\varepsilon$ perturbs the inferred action away from the deterministic best-response at each reasoning step (see \autoref{sec:methods} for details).
As a result, the outcome of the reasoning process of an agent may systematically deviate from what is expected from their reasoning level $k$, basing their strategic choices on a subjective interpretation of reality, possibly triggered by both external and internal factors such as the temptation for higher future rewards, as well as the anticipated behaviour of other individuals in the population.

We therefore relax the underlying assumption in \cite{lenaerts_evolution_2024}, wherein every individual shared the same reasoning process and made mistakes in the same way.
In this regard, our aim is to answer the following series of questions: \\
1) Under which conditions is rational behaviour replaced by different reasoning strategies; \\
2) Whether cognitively biased forms of reasoning are preferred to unbiased forms (with biases representing different types of systematic mistakes within each reasoning process);
and \\
3) Whether an evolutionary robust strategy emerges or whether multiple reasoning strategies may co-exist in the population.       


As in~\cite{lenaerts_evolution_2024}, the current analysis is also performed within the context of the Centipede Game~\cite{rosenthal_games_1981}, a sequential game with complete and perfect information that involves two individuals\footnote{For extensions to 3-player Centipede Games, see~\cite{rapoport_equilibrium_2003, murphy_population_2004}.} who take turns, over a total of $L$ steps, deciding between two actions ($T=$\textit{Take}, $P=$\textit{Pass}) regarding the split of a resource.
Player 1 (Player 2) can play $T$ at even (odd) steps.
Playing $T$ at a given step means ending the game and receiving a larger share of the resource than the co-player.
Playing $P$ means letting the other player decide what to do in the next step, unless the last decision node is reached, where Player 2 has to decide between two different splits (see \autoref{fig:centipede} for an example).

The payoff structure of a Centipede Game is designed such that solving the game via the game-theoretical method of backward induction leads a rational self-interested individual to play $T$ at every opportunity.
Hence, in the unique subgame-perfect Nash equilibrium (SPE), each player chooses to end the game as early as possible\footnote{Remarkably, all Nash equilibria in a Centipede Game require Player 1 to end the game at the first step.}.
Nevertheless, several studies have shown that this equilibrium strategy is rarely observed during behavioural experiments, especially when the resource to be shared grows with each step of the game~\cite{mckelvey_experimental_1992, fey_experimental_1996, palacios-huerta_field_2009}.
Here we focus on the \textit{Incremental Centipede Game} (ICG), where the resource starts at 0.5 and doubles at every step ~\cite{mckelvey_experimental_1992, rand_evolutionary_2012}.
In \autoref{fig:centipede}, we visualize the extensive form and the payoff structure for the case where $L=6$.

Several explanations have been proposed to bridge the gap between theoretical and experimental results in the Centipede Game. 
Typically, they can be categorized into two main groups, focusing either on insufficient cognitive ability to perform backward induction reasoning~\cite{mckelvey_quantal_1995, mckelvey_quantal_1998, stahl_evolution_1993, kawagoe_level-k_2012}, or on other-regarding motives that might interfere with self-interested decision making~\cite{bela_altruism_2022, krockow_exploring_2016, gamba_preferences-dependent_2015}.
Here, we argue that the potential evolutionary advantage of certain cognitive biases might have interfered with the development of backward induction reasoning, as these deviations might have been the result of heuristics adapted to deal with the complexity of uncertain environments~\cite{fawcett_evolution_2014, tversky_judgment_1974}.


Our contributions are threefold.
First, we show that a reasoning strategy with a systematic inference bias towards higher but uncertain rewards is favoured and becomes dominant under strong selection, whereas rational behaviour undergoes extinction.
Individuals employing such a biased inference process think, at each step, that higher-payoff outcomes are more likely to occur compared to individuals using the analogous unbiased strategy.
In this sense, their reasoning process, if compared to the unbiased reasoning type, systematically overestimates the probability that the ICG will end at a later node.
As a result, they tend (with some probability that depends on the noise parameter $\varepsilon$) to direct their decisions to stop the game at later steps with respect to an unbiased level--$k$ process, with no guarantee of actually reaching them.
Given these features, we label this reasoning strategy as a \emph{positively biased} reasoning strategy, whereas its inverse (i.e., directing their decisions to stop the ICG at earlier steps with respect to the unbiased reasoning type) is a \emph{negatively biased} reasoning strategy.

Our second contribution is to show that the sophistication of each individual, i.e., the depth of their reasoning processes, remains limited, underlining once more that explicit costs are not required to observe the emergence of bounded rationality~\cite{de_weerd_how_2013, lenaerts_evolution_2024, devaine_theory_2014}. 
Interestingly, we reveal the possibility of a co-existence between a myopic payoff-maximising strategy (i.e., ending the game in accordance with the personal highest payoff without any reasoning about the possible decision of the other player) and one level of positively biased reasoning (i.e., an individual with $k=1$ reasoning capacity whose mistakes in reasoning are biased towards higher future gains).
This finding is aligned with previous econometric analyses \cite{mckelvey_experimental_1992, kawagoe_level-k_2012}, which assumed the presence of a minority of altruistic players to align with experimental data. Yet we propose here that some individuals are driven by a desire to obtain the maximum payoff in the game rather than by a high pro-social orientation, thereby not relying on explanations related to other-regarding motives.

Finally, we identify that the length of the ICG, defined by the parameter $L$, is instrumental for the evolutionary success of positively biased reasoning, as longer exchanges lead to exponentially higher rewards. 
Specifically, higher values of $L$ amplify the range of cognitive noise $\varepsilon$ for which positively biased reasoning is the most frequent strategy in the population.
We observe nevertheless a limit to the window of noise values wherein this reasoning strategy is successful, as excessive cognitive noise renders any type of reasoning evolutionary detrimental.

Through these contributions, the present work further explains the emergence of the non-rational behaviour that is frequently observed in the experimental literature involving this sequential dilemma~\cite{krockow_exploring_2016} as well as others \cite{basu_travelers_1994, goeree_stochastic_1999}.
Our theoretical model, applied in the context of reciprocal exchanges of resources, thus corroborates the idea that certain cognitive biases, despite leading to systematic deviations from a rational judgment, can constitute an adaptive feature to successfully interact with our peers in this kind of social dilemmas.

\section{Results and Discussion}

To understand which form of reasoning, either unbiased or biased, is preferred by evolution, we examine the evolutionary dynamics of five types of reasoning.
We consider two basic types, namely the rational strategy of stopping the game as early as possible (i.e., the sub-game perfect equilibrium strategy, or \textit{SPE}), and the myopic payoff-maximizers who ignore the possible choices of the other player and stop the game at the node which maximises their personal reward (i.e., no-reasoning, \textit{NR}, or equivalently $k=0$).
In our case, the latter strategy corresponds to always playing \textit{Pass} for Player 1 and playing \textit{Take} at the last decision node for Player 2 (see \autoref{tab:backwardind}).
Additionally, we consider an unbiased (level--$k$) reasoning type that recursively applies ($k$ times) a best response perturbed by uniform cognitive noise over the player's action space, along with two biased (level--$k$) reasoning types.
The biased reasoning types are implemented similarly to the unbiased one, but the cognitive noise now directs the individual to end the game later or earlier than would be obtained by noiseless level--$k$ reasoning in the ICG.
As explained in the Introduction, the former will be referred to as a \emph{positively biased} reasoning type, since each step of the reasoning process is biased by a desire for future outcomes with a higher payoff (corresponding to later steps in the ICG) compared to what would be calculated by noiseless level--$k$ reasoning. 
In opposition, this makes the latter a \emph{negatively biased} reasoning type, as it is preferred to take before the step calculated by noiseless level--$k$ reasoning.  
Details are provided in \autoref{sec:methods}.

\subsection*{Positively biased reasoning co-evolves with bounded rationality.}

\autoref{fig:effect-of-beta} answers our first question: 
we show the outcome of the co-evolution between reasoning types and $k$-levels as dictated by the stationary distribution of the stochastic evolutionary model for different values of the selection strength $\beta$ (see \autoref{sec:methods}) ~\cite{rand_evolutionary_2012, lenaerts_evolution_2024}. 
To calibrate the output of our model, we determine the relevant level of cognitive noise $\varepsilon$ comparing our predictions with the experimental data from Kawagoe and Takizawa \cite{kawagoe_level-k_2012}. 
The best fitting is provided for $\varepsilon \approx 0.19$, as shown in \autoref{fig:fitting} and \autoref{sec:appendixA}. 
For comparison purposes, the same analysis and fitting are performed with the data on the same game from the seminal work by McKelvey and Palfrey \cite{mckelvey_experimental_1992}, producing $\varepsilon \approx 0.06$ (see \autoref{sec:appendixA} for further details). 

Panel \textbf{A} shows the abundance of reasoning types as given by the stationary distribution for different values of selection strength $\beta$ and the calibrated cognitive noise $\varepsilon \approx 0.19$. 
The aim here is to understand how increasing the selective pressure (i.e., higher $\beta$) affects the likelihood of ending up with any of the reasoning strategies studied here. 
When $\beta\simeq 0$, the evolutionary dynamics correspond to the process of neutral drift, thus each reasoning strategy is almost equally present in the population.  
As $\beta$ grows, the payoff differences between strategies become increasingly relevant. 
As can be observed, the positively biased reasoning type takes over the population, fully dominating when $\beta \geq 0.4$.    
This preference for positively biased reasoning grows together with a preference for performing $k=1$ reasoning steps, as can be seen in Panel \textbf{B}. 
In fact, after a minor peak in the frequency of $k=0$ individuals, the $k=1$ type eventually takes over the entire population. 

The right panels in \autoref{fig:effect-of-beta} show the results for the same model fitted to the experimental data in \cite{mckelvey_experimental_1992}.
With a lower value of $\varepsilon$ (i.e., $\varepsilon\approx 0.06$), the mistakes within the reasoning processes are less frequent than in the previous case.
The differences between reasoning types are therefore not as marked.
However,  positively biased reasoning is again the most frequent type under strong selection ($\beta\ge1$), being adopted by more than 60\% of the population (Panel \textbf{C}).
Moreover, the increased level of reasoning accuracy allows for the development of a higher reasoning capacity, with $k=2$ now being the most frequent level of reasoning, adopted by 60 to 68\% of the population under strong selection, followed by $k=1$ and $k=3$, both with comparable frequencies (Panel \textbf{D}). 

In the following section, we examine the stochastic evolutionary dynamic that leads to this result in more detail. 
Before, we further observe that as the stochasticity of the imitation process decreases (i.e., for larger values of $\beta$), higher levels of reasoning ($k=4$ and $k=5$) progressively disappear from the population, meaning that the sophisticated individuals are replaced by the less sophisticated ones.
This result is in agreement with both theoretical~\cite{de_weerd_how_2013, lenaerts_evolution_2024} and experimental~\cite{camerer_cognitive_2004, kawagoe_level-k_2012} findings.
Notably, higher levels of reasoning do not need to be explicitly penalized through the introduction of any computational cost because every additional backward step in the reasoning process has the implicit cost of propagating and amplifying the effect of cognitive noise. 
Therefore, the population eventually finds the best trade-off between sophistication and propagation of errors, leading to the presence of intermediate reasoning levels $k=1,2,3$ when $\varepsilon$ is sufficiently small and to the invasion of the level $k=1$ at higher $\varepsilon$.


Remarkably, the pure strategy associated with the \textit{SPE} (i.e., to stop the game as early as possible) never manages to invade the others.
This novel outcome is at odds with a previous evolutionary model proposed by Rand and Nowak~\cite{rand_evolutionary_2012} which showed that this strategy would eventually prevail under strong selection when the population only plays pure strategies. 
Our model avoids this outcome because each strategy is a noisy introspective process~\cite{goeree_stochastic_1999, goeree_model_2004}, where iterated conjectures about others' decisions might undergo stochastic deviations from the deterministic rational predictions.
This result therefore underlines once again the importance of the introduction of cognitive noise, and therefore of the generated mixed strategies, to obtain a more flexible and insightful model.

\subsection*{A co-existence dynamic emerges from biased reasoning.}

In \autoref{fig:diagram}, we perform a more detailed analysis of the evolutionary dynamics to explain the previous observations.
Panel \textbf{A} shows the invasion diagram (i.e. the reduced Markov chain connecting all possible monomorphic population states) for the calibrated values $\varepsilon=0.19$ and $\beta=0.063$, corresponding to the settings indicated with the white dotted lines in \autoref{fig:effect-of-beta}A and B. 
For a better visualization, the strategies are restricted to $k\leq 2$, as higher $k$-levels cover a negligible portion of the stationary distribution.
The invasion diagram is a useful tool to understand the dominance relationships between strategies under the small-mutation limit approximation. A directed edge from node $i$ to node $j$ in this diagram, indicates that a single mutant adopting strategy $j$ is able to fixate in a population of $i$s with a probability higher than random drift \cite{sigmund_calculus_2010}. An Evolutionary Robust Strategy (ERS) is a strategy with no outgoing arrows ~\cite{stewart_extortion_2013}.

As we can see for these parameter settings, there does not exist any ERS\footnote{As mentioned previously, when $\varepsilon\approx 0.19$ the  positively biased reasoning type with $k=1$ turns into an ERS for $\beta > 0.4$}. 
Instead, an interesting cycle is present, involving the three most frequent strategies, namely the  positively biased reasoning strategy with $k=1$, $B^{+}(1)$, the no-reasoning (\textit{NR}) or payoff-maximising strategy, and $k=1$ level of unbiased reasoning $U(1)$. 
What is interesting is that the arrow connecting $B^{+}(1)$ and \textit{NR} is pointing in both directions, which may indicate a potential for co-existence of both reasoning strategies~\cite{auger_hawk-dove_1998}.

This hypothesis is confirmed in Panel \textbf{B}, where we consider the replicator dynamic \cite{schuster1983replicator,cressman2014replicator} restricted to the three strategies \textit{NR}, $B^{+}(1)$ and $U(1)$, in which the population spends approximately 80\% of the time according to the stationary distribution visualised in Panel \textbf{A}. 
The results of the replicator dynamic are visualised through the simplex covering all population configurations for the three aforementioned strategies, including the anticipated dynamics in each point of the simplex. 
One can clearly observe that a stable co-existence point appears for the calibrated cognitive noise $\varepsilon\approx0.19$, wherein 60\% of the population uses level-1  positively biased reasoning ($B^{+}(1)$), while the remaining 40\% uses no reasoning (\textit{NR}). At each interior point of the simplex, the replicator dynamic is directed towards this novel equilibrium.  
We thus see that the game and its outcomes are transformed due to the presence of biased and unbiased reasoning strategies~\cite{lenaerts_evolution_2024,nowak2006five,taylor2007transforming}, revealing the impact that (biased) reasoning strategies may have on decision-making.

This transformation of the game and resulting co-existence dynamic is further explored in Panels \textbf{C} and \textbf{D} of \autoref{fig:diagram} where the complete Markov chain connecting the monomorphic population with \textit{NR} to the one containing only $B^{+}(1)$ is analysed. 
Both plots show the interplay between the intensity of cognitive noise $\varepsilon$ and the co-existence point when the game is restricted to the two aforementioned strategies.
As we can see, when $\varepsilon$ is sufficiently small ($\varepsilon\approx 0.1$), strong selection leads to the full invasion of individuals adopting the strategy $B^{+}(1)$.
Larger values of $\varepsilon$ allow the emergence of a co-existence point, where the fractions of the \textit{NR} type increases with $\varepsilon$, until the amount of cognitive errors is so large that performing any reasoning becomes evolutionary detrimental.
The results remain robust when studying the full Markov chain under mutations (see Panel \textbf{D}).

Overall, these results confirm the importance of the  positively biased reasoning strategy within the context of the ICG.
Indeed, the possibility of developing positively biased reasoning allows the population to enter a new stable equilibrium, where the majority of the population is made of positively biased reasoners that co-exist with a minority of myopic payoff-maximizers (assuming a sufficiently small probability of reasoning mistakes).
This novel equilibrium, while leading to a better agreement with the experimental reference, is also aligned with previous theoretical and behavioural studies arguing that humans' sophistication in a population is often heterogeneous and that ``naive" types can survive and co-exist with more sophisticated individuals~\cite{heller_three_2015, camerer_cognitive_2004, crawford_lying_2003}.

\subsection*{Longer exchanges promote the emergence of positively biased reasoning.}

\begin{figure}[!ht]
\centering
\resizebox{\textwidth}{!}{%
\includegraphics{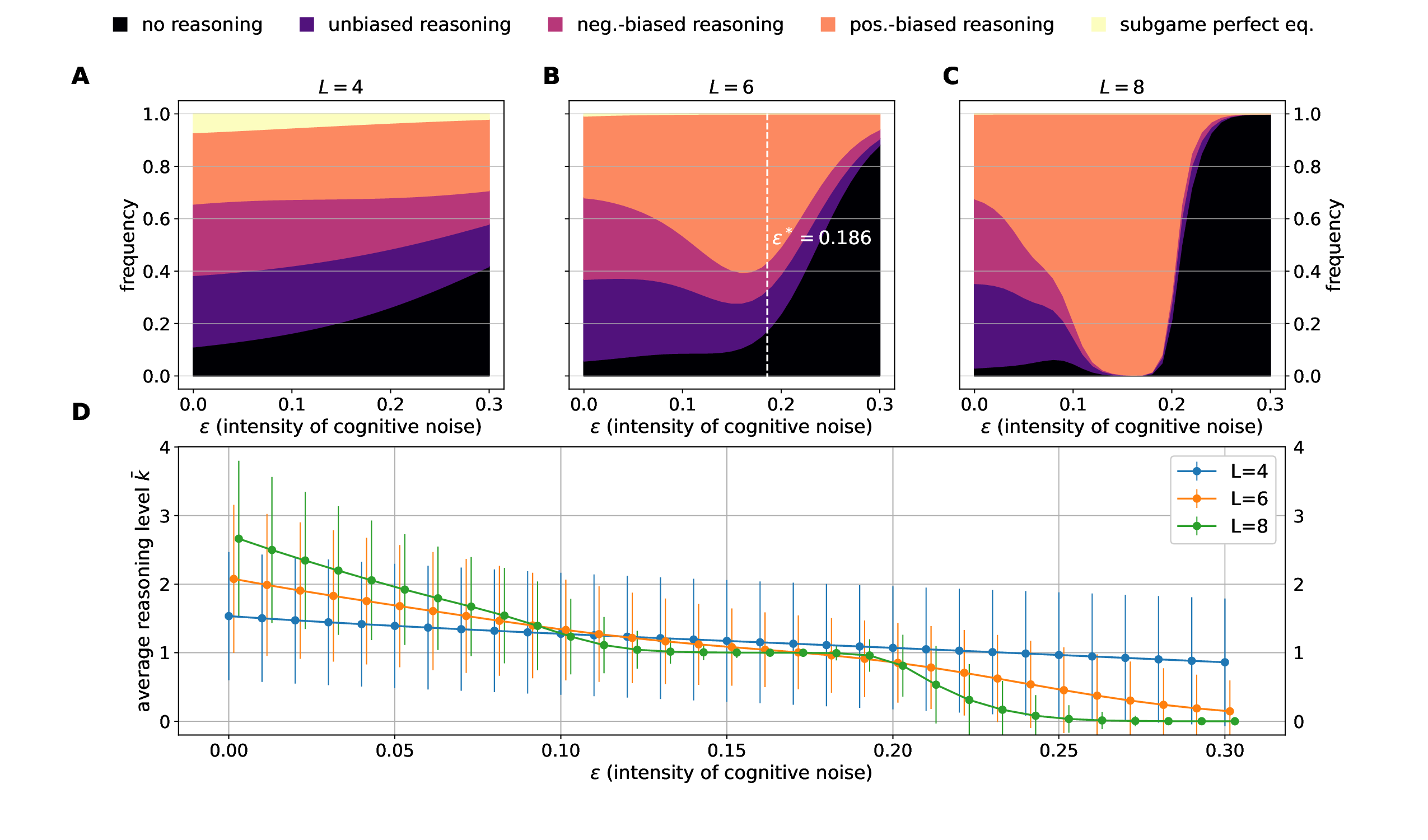}
}
\caption{Longer exchanges promote the emergence of positively biased reasoning.
The effect of the number of steps of the ICG, represented by the parameter $L\in\left\{4,6,8\right\}$, on the evolution of reasoning kernels (panels \textbf{A-C}) and reasoning levels (panel \textbf{D}) is shown for variable cognitive noise $\varepsilon$.
For each value of $\varepsilon$, we compute the stationary distribution $\phi$ of our dynamical system under the assumption of rare mutations. 
In panels \textbf{A-C}, the frequencies of each strategy are aggregated by reasoning kernel, $f_{ker}=\sum_{k=1}^{L}\phi_{s_{ker, k}}$ for $ker\in \left\{U, B^{+}, B^{-}\right\}$.
For completeness, the figures also include the frequency of the pure strategy associated with no reasoning (i.e., reasoning level $k=0$) and the sub-game perfect equilibrium (SPE) strategy of stopping the game as early as possible.
In panel \textbf{D}, we report the average reasoning level of the population, 
$\bar{k}=\sum_{k=1}^{L}k f_k$ where $f_{k}=\sum_{ker\in \left\{U, B^{+}, B^{-}\right\}}\phi_{s_{ker, k}}$, along with its standard deviation plotted in form of error bars.
The white dotted line in panel \textbf{B} corresponds to the value of cognitive noise $\varepsilon^*$ of optimal fitting with the data in~\cite{kawagoe_level-k_2012}. 
The population size is fixed to 100 individuals, and the strength of selection is $\beta^*=0.063$.
}
\label{fig:effect-of-eps}
\end{figure}

The length of the ICG, represented here by the parameter $L$, is a key factor to secure the evolutionary success of positively biased reasoning.
Indeed, longer ICGs translate into a much larger resource to be shared among the players because of the exponential growth inherent in the payoff structure.
In \autoref{fig:effect-of-eps} we explore the effect of the intensity of cognitive noise $\varepsilon$ on the stationary distribution of our model when the selection strength $\beta$ is fixed at 0.063, for three values of $L$ ($L=4, 6, 8$ in Panels \textbf{A,B,C}, respectively).

When $L=4$, the difference in frequency between the unbiased,  positively biased, and  negatively biased reasoning is negligible, regardless of the intensity of cognitive noise. 
When $L=6$, the incentive to end the game at later nodes is exponentially higher, leading to a prevalence of positively biased reasoning for a window of $\varepsilon$ values:
in fact, more than half of the population develops a positivity bias within the range $0.11<\varepsilon<0.2$.
Finally, when $L=8$, the same effect is even more pronounced: now, the region where more than half of the population chooses positively biased reasoning extends to the range $0.05<\varepsilon<0.2$, and in the subset of values $0.14<\varepsilon<0.17$ the entire population adopts the strategy previously mentioned $B^{+}(1)$.

In each of the three scenarios, however, a larger probability of cognitive errors renders the reasoning process increasingly harmful, regardless of the reasoning process employed.
This is particularly evident in the plot showing the decay of the average reasoning level $\bar{k}$ (\autoref{fig:effect-of-eps}\textbf{D}). 
The figure reports the mean and the standard deviation of the distribution of reasoning levels within the population in the stationary state in function of the cognitive noise. 
We can see how, as $\varepsilon$ increases, the myopic payoff-maximizers progressively spreads across the population. This effect is more pronounced as the length of the ICG increases to $L=8$.

Once again, the presence of the strategy associated with the \textit{SPE} is negligible, regardless of the value of $\varepsilon$.
This result is remarkable because this pure strategy, despite being deterministic, is indirectly affected by the amount of mistakes committed within the population of interacting agents:
In a population of sophisticated individuals who, nonetheless, make reasoning mistakes, it is better to take advantage of these mistakes and opt for a myopic maximisation of one's own payoff, rather than performing the rational reasoning process of backward induction.

\section{Conclusions}
We presented a model based on the ICG to study the co-evolution of strategic reasoning with and without two cognitive biases, which were referred to as positively  and negatively biased reasoning, in the framework of Evolutionary Game Theory.
We showed how evolutionary dynamics in finite populations can lead boundedly rational players in this game to develop a positivity bias while recursively reasoning about the potential move of their opponents.
Furthermore, the emergence of positively biased reasoning in this sequential setting appears to be strongly linked to two factors, namely the presence of individuals acting as myopic payoff maximisers, and the promise of higher future gains (captured here by the length of the ICG and the exponential growth of the resource to be shared among the players). 


While our work was restricted to an evolutionary analysis of positively  and negatively biased reasoning strategies in the ICG, the model may be interpreted to reflect specific cognitive and psychological phenomena.
As an example, the aforementioned reasoning strategies could be potentially associated with \emph{wishful thinking}~\cite{aue_neural_2012,neumann_bounded_2014, yildiz_wishful_2007} and its counterpart \emph{defensive pessimism}~\cite{vosgerau_how_2010, sanna_defensive_1998, bateson_agitated_2011}.
Wishful thinking is described in the literature as a distortion of cognitive forecasting processes towards more optimistic expectations, and it implies the overestimation of the likelihood of desirable events.
This concept therefore fits, to a certain extent, our model, as the ICG formalizes the partition of a growing resource between parties that hold beliefs about each other and have desires for possibly different outcomes.

Wishful thinking, however, is an extremely broad phenomenon. 
Moreover, the related experimental literature exploring the link between wishful thinking and strategic decision-making is rather limited and produced mixed results (see e.g.,~\cite{krizan_wishful_2009, krizan_influence_2007} for a review).
It is therefore not obvious that wishful thinking is the source of the behavioural deviations presented here or that the model truly captures the intricacies of this concept.
Alternative interpretations, such as a risk-taking tendency~\cite{dekel1999evolution,brocas2025children}, may also align with the results of our model.  
A more comprehensive analysis of the link between the model and these cognitive interpretations is therefore needed, which is left for future work.

Our study was also limited to the investigation of cognitive biases in the form of consistent deviations from a rational judgment of the opponent's future choice.
The cognitive bias of each player was unequivocally translated into a specific behaviour.
In this sense, our study assumed that the relationship between behavioural strategies and cognitive biases is unambiguous, while each strategy could, in reality, be the outcome of multiple cognitive processes and heuristics~\cite{fawcett_evolution_2014}, as remarked above.
Moreover, our model assumed a static belief framework in which individuals hold fixed prior beliefs about their co-players. It also imposes that individuals share the same beliefs throughout the population and that all players use the same type of recursive reasoning process, where opponents always possess a comparatively lower capacity for reasoning~\cite{lenaerts_evolution_2024}. 
Future work can relax these assumptions to examine how robust the results remain while also opening the door to alternative models to explore other aspects of cognitive biases and related psychological phenomena that were not yet considered here.

To conclude, the insights provided by our model, while confirming previous experimental~\cite{johnson_evolution_2011, kawagoe_level-k_2012} and theoretical~\cite{lenaerts_evolution_2024, de_weerd_how_2013, neumann_bounded_2014} results, show that the deviations from the rational behaviour prescribed by game theory can be the outcome of evolutionary processes in which reasoning heuristics were developed in order to better interact in complex and uncertain environments.
In particular, positively biased reasoning and bounded rationality proved to be adaptive features in the context of reciprocal exchanges of a shared resource that grows over time, in line with theories of the evolution of human social cognition claiming that we developed these non-rational behaviours by means of our remarkable ability to understand the intentionality of our peers.

\section{Methods}\label{sec:methods}

\subsection{Notation}
Let us first introduce the notation that we will use throughout this section:
\begin{itemize}
    \item The subscript $i\in\left\{1,2\right\}$ denotes the role of the player, while the classical notation $-i$ denotes the role of the co-player;
    \item $\mathcal{A}_i$ represents the set of possible actions of the player in role $i$: specifically, $\mathcal{A}_1=\left\{0,2,\ldots,L-2, L\right\}$ and $\mathcal{A}_2=\left\{1,3,\ldots,L-1, L\right\}$ with $|\mathcal{A}_1|=|\mathcal{A}_2|=1 + L/2$ (note that the action $L$, shared by both players, means to always play \textit{Pass});
    \item $T_i$ $\in\mathcal{A}_i$ denotes the step at which the player in role $i$ will end the game; 
    \item $\sigma_i$ represents the probability distribution over the different steps at which the player in role $i$ may end the game.
\end{itemize}

\subsection{Strategies}\label{sec:strategies}

Given an ICG of $L$ steps, our game-theoretical model considers two pure strategies and $3 \times (L-1)$ generative strategies\footnote{We use the term \emph{generative} as opposed to \emph{mixed} because in each interaction, these strategies need to generate the action based on a noisy level--$k$ recursive process. While they implicitly encode a mixed strategy, i.e. a probability distribution over the action space of both players, they are defined by the reasoning level $k>0$ and one of the three reasoning types, as it is further explained in Methods.}.
The pure strategies consist of the sub-game perfect Nash equilibrium (SPE) strategy and a myopic payoff maximisation strategy.
The former will decide to end the game as early as possible (in line with full backward induction).
The latter aims at acquiring the highest possible personal reward, ignoring the potential choices of the co-player.
If $L=6$, for instance, when acting as Player 1, this strategy will play $T_1=6$ because $\pi_1(6)=25.6$ is the highest payoff of the game for Player 1 (see \autoref{fig:centipede}), without taking into account the other player's likely \textit{Take} move at the fifth node. Likewise, a myopic Player 2 would stop the game at the step $T_2=5$ to obtain the maximum payoff $\pi_2(5)=12.8$, without considering that Player 1 would likely stop the game at the preceding node (see \autoref{fig:centipede} and \autoref{tab:backwardind}).

The $3 \times (L-1)$ generative strategies are defined using the level--$k$ framework ~\cite{stahl_evolution_1993, lenaerts_evolution_2024}. 
Each generative strategy is determined by a value for $k$ and the type of (un)biased reasoning they will use to arrive at a decision.
Each reasoning level $k$ corresponds to $k$ steps of backward induction reasoning, where players might deviate from the rational best response at each reasoning step due to cognitive noise, defined by a parameter $\varepsilon\in\left[0, 1\right]$.
At $k=0$, which corresponds to no-reasoning (NR), the strategy corresponds to the myopic payoff-maximizer explained earlier.
When $k>0$, level--$k$ individuals perform a recursive reasoning process assuming that the co-player belongs to the type $k-1$.

In a situation where individuals do not make any reasoning mistakes ($\varepsilon=0$), this process can be formulated as follows:
\begin{align}\label{eq:levk}
    T_i(k) = \begin{cases}
        \argmax_{0\le t\le L}\pi_i(t), & \text{if } k = 0 \\
        \BR_i(T_{-i}(k-1)), & \text{otherwise}
    \end{cases}
\end{align}
where $\BR_i(\cdot)$ encodes the best response of Player $i$ given the action of Player $-i$.
This case is essentially analogous to ~\cite{rand_evolutionary_2012}, with the reasoning level $ k=L-1$ being equivalent to the SPE strategy (see  \autoref{tab:backwardind}).

Since we are interested in the evolution of cognitive biases and strategic reasoning, we focus on the case $\varepsilon>0$:
in particular, at each reasoning step, we let players deviate from the best-response with probability $\varepsilon\in[0,1]$~\cite{lenaerts_evolution_2024}, i.e., Player $i$ will now choose their action according to a noisy best response, $\NBR_i(\cdot)$, which replaces $\BR_i(\cdot)$ in \autoref{eq:levk}.
Resuming the example mentioned above, if Player 2 now performs one reasoning step ($k=1$), they will take into account that a \textit{fictional} Player 1 with $k=0$ follows the pure strategy $T_1(0)=6$, so Player 2 will adopt strategy $T_2(1)=5$ with probability $1-\varepsilon$, but will deviate from it with probability $\varepsilon$. 
Depending on the reasoning type, these deviations can occur uniformly on the whole action space if the reasoning process is unbiased, or they can be skewed towards earlier or later steps in case of negatively or positively biased reasoning, respectively.

Each reasoning process will thus produce a role-dependent probability distribution over the actions of each player, $\sigma_1(k)$ and $\sigma_2(k)$, corresponding to the probability distributions over the action spaces of Player 1 and Player 2, respectively:
\begin{align}
    \sigma_1(k) = \left[\mathbb{P}(T_1(k)=0),\ \mathbb{P}(T_1(k)=2),\  \ldots,\ \mathbb{P}(T_1(k)=L)\right],\notag\\ 
    \sigma_2(k) = \left[\mathbb{P}(T_2(k)=1),\ \mathbb{P}(T_2(k)=3),\  \ldots,\ \mathbb{P}(T_2(k)=L)\right].\notag
\end{align}
The recursive reasoning in \autoref{eq:levk} then corresponds to an iterative application of the law of total probability, where each element of the vectors $\sigma_1(k)$ and $\sigma_2(k)$ equals:
\begin{eqnarray}\label{eq:totalprob}
    &\mathbb{P}(T_i(k)=t_i) =\\
    &\sum_{t_{-i}\in\mathcal{A}_{-i}} \mathbb{P}(\NBR_i(t_{-i})=t_i\ |\ T_{-i}(k-1)=t_{-i}) \cdot \mathbb{P}(T_{-i}(k-1)=t_{-i})\notag
\end{eqnarray}
with $i\in\left\{1,2\right\}$, $k>0$, and $t_i\in\mathcal{A}_i$.
The overall probability of Player $i$ playing action $t_i$ is thus the sum of the probabilities that $t_i$ is chosen as the response to a fictional level--$(k-1)$ opponent playing the action $t_{-i}$, over all possible actions in the set $\mathcal{A}_{-i}$.

It can be shown that the strategy of a player covering both roles with equal probability can be computed through the following formula (see \autoref{sec:appendixB} for further details):
\begin{eqnarray}
    \sigma(k) = \sigma(0)\cdot M(\varepsilon)^{k},
\end{eqnarray}
where $\sigma(k)$ with $0\leq k \leq L-1$ is the concatenation of the two role-dependent strategy vectors $\sigma_{1}(k)$ and $\sigma_{2}(k)$, and $M(\varepsilon)$ is a matrix which depends on the intensity of the cognitive noise $\varepsilon$ and encapsulates the noisy best responses for both roles.

Each strategy in our model is therefore uniquely determined by three components of the reasoning process: 
(i) the starting point, $\sigma(0)$; (ii) the reasoning kernel, $M(\varepsilon)$; and (iii) the depth of the reasoning process, $0\le k < L$. 
This framework has an intuitive and straightforward interpretation: 
each individual has a \textit{prior behaviour}, $\sigma(0)$, about the action to choose without any reasoning, and they modify it by applying the reasoning kernel $M$ (representative of their \textit{mind}) iteratively for $k$ times.
As stated previously, we fix the prior belief $\sigma(0)$ of each individual such that $k=0$ corresponds to an agent who myopically aims at maximising the payoff of the game, leading to $\sigma(0)=\left[0,\ 0,\ 0,\ 1,\ 0,\ 0,\ 1,\ 0\right]$ for $L=6$.

To study the evolution of cognitive biases, we propose the following reasoning kernels:
\begin{itemize}
    \item an \textit{unbiased} reasoning kernel, $M_{U}(\varepsilon)$, where the best response is chosen at each step with probability $1-\varepsilon$ while any other action is chosen with uniform probability $2\varepsilon/L$;
    \item a \textit{negatively biased} reasoning kernel, $M_{B^{-}}(\varepsilon)$, where deviations from the best response only happen to shift towards earlier nodes of the game, i.e., lower-payoff outcomes;
    \item a \textit{positively biased} reasoning kernel, $M_{B^{+}}(\varepsilon)$, where deviations from the best response only happen in direction of later nodes, i.e., outcomes where payoffs are higher but uncertain.
\end{itemize}
We denote the level--$k$ strategies of unbiased, positively biased, and negatively biased reasoning by $U(k)$, $B^{+}(k)$, and $B^{-}(k)$, respectively.
In \autoref{sec:appendixB} we report the three matrices $M_{U}(\varepsilon)$, $M_{B^{+}}(\varepsilon)$, and $M_{B^{-}}(\varepsilon)$ in the case where $L=6$.


\subsection{Payoffs}
We now describe the computation of the payoffs resulting from the interaction between two strategies.
Let us suppose two players, Alice and Bob, adopt strategies $\sigma_A$ and $\sigma_B$ respectively. 
As stated previously, each of these vectors consists of a pair of role-dependent probability distributions, namely $(\sigma_{1A},\sigma_{2A})$ and $(\sigma_{1B},\sigma_{2B})$.

Here, we assume that Alice and Bob play against each other covering both roles with equal probability, so that the game becomes symmetric~\cite{rand_evolutionary_2012}. 
This means that, for instance, the expected payoff of Alice, $\Pi_A(\sigma_A, \sigma_B)$, is given by the average between the expected payoff she would obtain while playing as Player 1 against Bob playing as Player 2 and the expected payoff in the opposite scenario. 
Thus, we obtain the following equation:

\begin{eqnarray}
    \Pi_A(\sigma_A, \sigma_B)
    =\frac{1}{2}\sum_{t=0}^{L}\left\{\pi_1(t)\cdot\mathbb{P}\left(T^{\min}_1=t\right) + \pi_2(t)\cdot\mathbb{P}\left(T^{\min}_2=t\right) \right\},
\end{eqnarray}
where $T^{\min}_1$ and $T^{\min}_2$ are the two random variables associated with the distribution of the terminal node of the game on the two admissible scenarios, where Alice (Bob) plays as Player 1 (2) and Player 2 (1), respectively:

\begin{align}
    T^{\min}_1 := \min\left\{\sigma_{1A}, \sigma_{2B}\right\},\ 
    T^{\min}_2 := \min\left\{\sigma_{2A}, \sigma_{1B}\right\}.\notag
\end{align}

Note that, since these two random variables are independent and defined over the same support, the cumulative distribution function of the minimum between the two can be computed analytically through the following equation:
\begin{eqnarray}
    \mathbb{P}(T^{\min}_i\le t)
    = 1 - \mathbb{P}(\sigma_{iA}> t) \mathbb{P}(\sigma_{-iB}> t).\notag
\end{eqnarray}

\subsection{Evolutionary dynamics}
As mentioned in the Introduction, we adopt an approach based on Evolutionary Game Theory to understand how cognitive biases and strategic reasoning might have co-evolved in the long term. 
In brief, this approach approximates the evolution of the behaviour of a population of $Z$ individuals interacting among each others according to their given strategies, via analytical methods or numerical simulations~\cite{fatima_learning_2024}. 
Borrowing ideas from evolutionary biology, a strategy can propagate within the population if it is associated with higher fitness, i.e., yield a higher expected payoff, than the others.

At each generation step, one individual $X$ is randomly chosen to update their strategy by either undergoing a stochastic mutation with probability $\mu$ or by imitating a better co-player~\cite{fudenberg_imitation_2006}.
In the latter case, another individual $Y$ is randomly sampled from the population to act as a potential role model, and will be imitated with probability $p=1/(1+\exp{(\beta(\Pi_X-\Pi_Y))})$, which increases with the fitness difference between $Y$ and $X$ ~\cite{traulsen_stochastic_2006}. 
The parameter $\beta$ represents the intensity of selection, i.e., the strength of imitation: 
when $\beta=0$, imitation occurs with 50\% chance, corresponding to the process of neutral drift; when $\beta$ is large, imitation becomes sensitive to the slightest fitness difference, leading to an almost deterministic evolutionary process.
The number of states of this birth-death process, however, quickly becomes intractable via analytical methods.

\begin{figure}[!ht]
\centering
\resizebox{\textwidth}{!}{%
\includegraphics{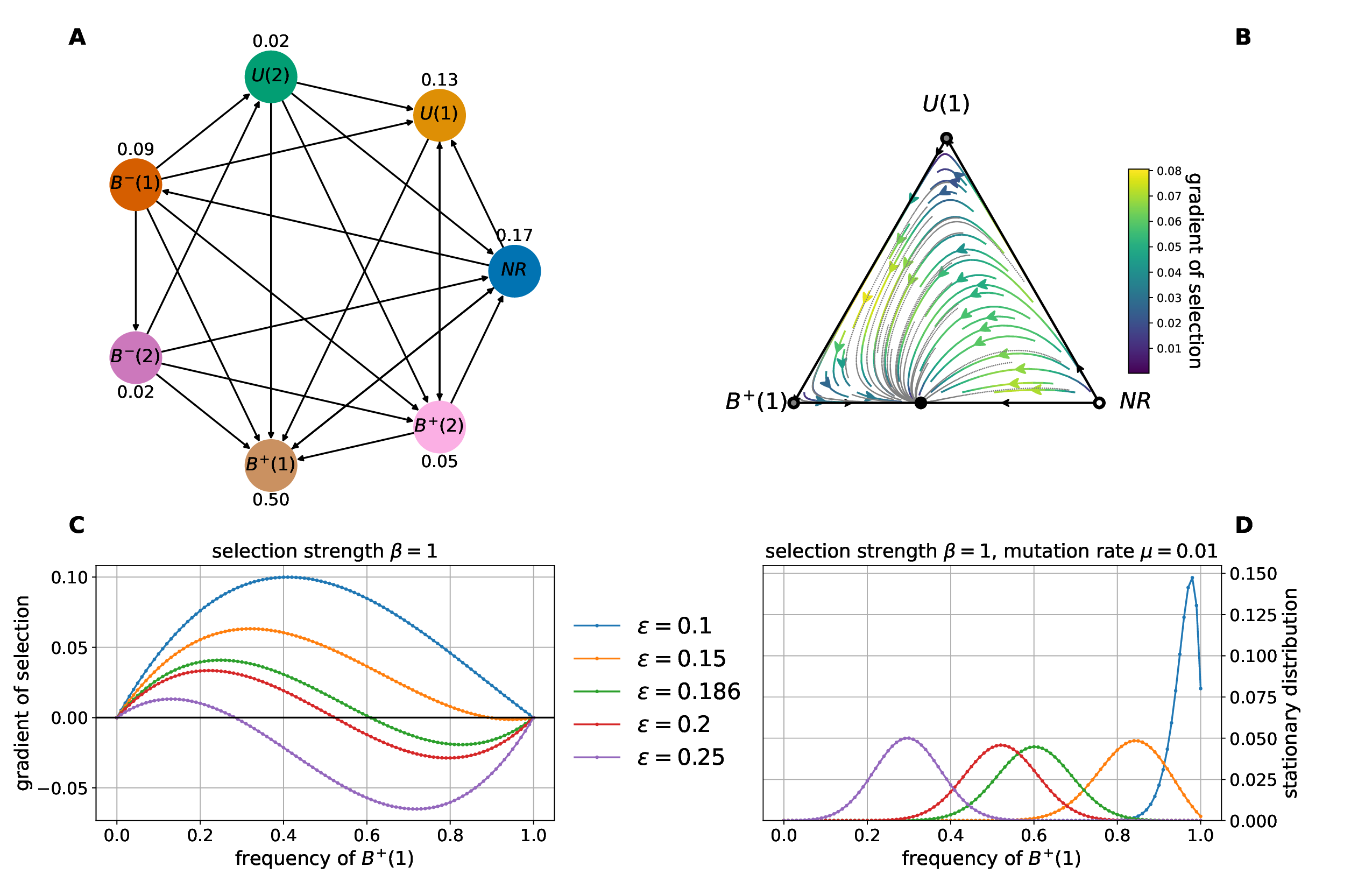}
}
\caption{No-reasoning and level--1 positively biased reasoning co-exist under limited cognitive noise.
Panel \textbf{A} shows the invasion diagram related to the evolutionary dynamics of our model under the small mutation limit, for $\beta=0.063$ and $\varepsilon=0.186$.
The strategies are limited to $k\leq 2$, and when $k>0$ the reasoning kernel can be unbiased ($U$), positively biased ($B^{+}$) or negatively biased ($B^{-}$).
The edges in the graph represent the transitions between different monomorphic states:
an outgoing edge from strategy $A$ to $B$ means that $A$ is invaded by $B$.
The number next to each node is the fraction of time spent in each monomorphic state at the stationary regime.
Panel \textbf{B} focuses on the evolutionary dynamics in an infinite population between the three most frequent strategies in Panel \textbf{A}, i.e., $B^{+}(1)$, \textit{NR}, and $U(1)$.
We plot the gradient of selection and the fixed points associated to the replicator equation.
The black and white circles represent stable and unstable equilibria respectively, whereas gray circles are saddle points. 
The arrows indicate the direction of the selective pressure.
Finally, Panels \textbf{C} and \textbf{D} show the effect of the cognitive noise $\varepsilon$ on the location of the co-existence point between the two strategies $B^{+}(1)$ and \textit{NR} under strong selection ($\beta=1$). 
The population size in Panels \textbf{A,C,D} is $Z=100$.
}
\label{fig:diagram}
\end{figure}

A possible solution is the so-called \textit{small-mutation limit}, i.e., the case where mutations are negligible ($\mu\to 0$)~\cite{hindersin_computation_2019}.
This approximation allows us to reduce the size of the Markov chain to the number of strategies of the game. 
Indeed, with this approximation, the time interval between two mutations is sufficiently large that evolution will lead to the fixation of one strategy in the population before the next mutation leads to the appearance of a new strategy. 
Thus, at any time, there will be at most two strategies simultaneously present in the population. 
The rare-mutation limit leads to an embedded Markov chain whose states correspond to the different homogeneous configurations of the population in which everyone plays the same strategy.
Most results in this work, except for Panels \textbf{B-D} of \autoref{fig:diagram}, are obtained under the small-mutation limit.

A further approximation, used in Panels \textbf{B} of \autoref{fig:diagram} is given by the limit of infinite populations. 
The dynamics in this case is represented by the replicator equation \cite{schuster1983replicator,cressman2014replicator}.
This differential equation expresses a deterministic selection process where the frequency of a type $i$, $x_{i}$, increases if it has higher fitness, $\Pi_{i}$, than the average fitness of the population $\overline{\Pi}$. 
In formulas, $\dot{x}_i = x_{i}(\Pi_{i}(x) - \overline{\Pi}(x))$.
When the gradient is 0, i.e., $\dot{x}_{i}=0$, we have an equilibrium point.
If the gradients in a small neighbourhood of an equilibrium $x^*$ point toward $x^*$, the equilibrium is said to be evolutionary stable. 
This definition means that an Evolutionary Stable State is robust to small changes in the population, i.e., a mutant will not drive the population to a different state.
This solution concept, more strict than the well known Nash equilibrium, allows to characterize complex evolutionary processes through relatively simple differential equations.
 
\subsection{Code availability}
The software implementation of the evolutionary processes is based on EGTTools\footnote{\href{https://egttools.readthedocs.io/en/latest/}{https://egttools.readthedocs.io/en/latest/}}~\cite{fernandez_domingos_egttools_2023}, an open-source hybrid C++/Python library that provides both analytical and numerical methods to study game-theoretical problems within the framework of Evolutionary Game Theory. The code to reproduce all the results presented in this work can be found at the following GitHub \href{https://github.com/MarcoSaponara/centipede-bias.git}{repository}\cite{marco_saponara.2025.15838520}.

\section{Acknowledgments}
The authors gratefully acknowledge the research support of the F.R.S-FNRS (project grant 40007793).
TL further acknowledges the support of the Service Public de Wallonie Recherche (grant 2010235–ARIAC) by DigitalWallonia4.ai and the Flemish Government through the AI Research Program.
EFD is supported by an F.W.O. Senior Postdoctoral Grant (12A7825N). 
The authors also thank Axel Abels and the anonymous reviewers for their useful comments and suggestions, which allowed us to improve this article.

\clearpage



\section{Figures \& Tables}
\subsection{Figures}

\begin{figure}[!ht]
\centering
\resizebox{\textwidth}{!}{%
\includegraphics{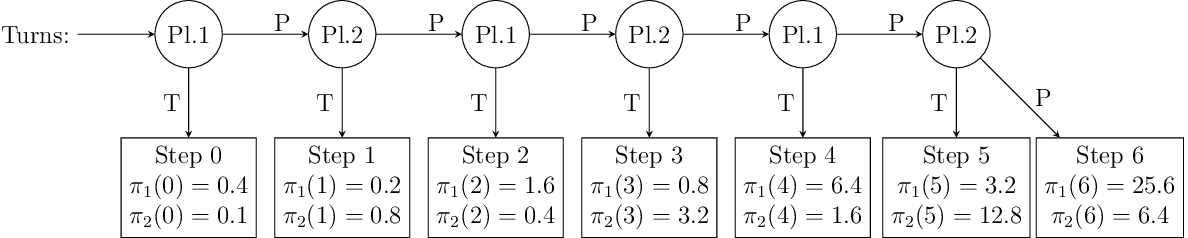}
}
\caption{Extensive form of the six-step Incremental Centipede Game with exponential growth. The notation $\pi_i(t)$ denotes the payoff Player $i$ would get if the game ends at Step $t$.}
\label{fig:centipede}
\end{figure}

\begin{figure}[!ht]
\centering
\resizebox{\textwidth}{!}{%
\includegraphics{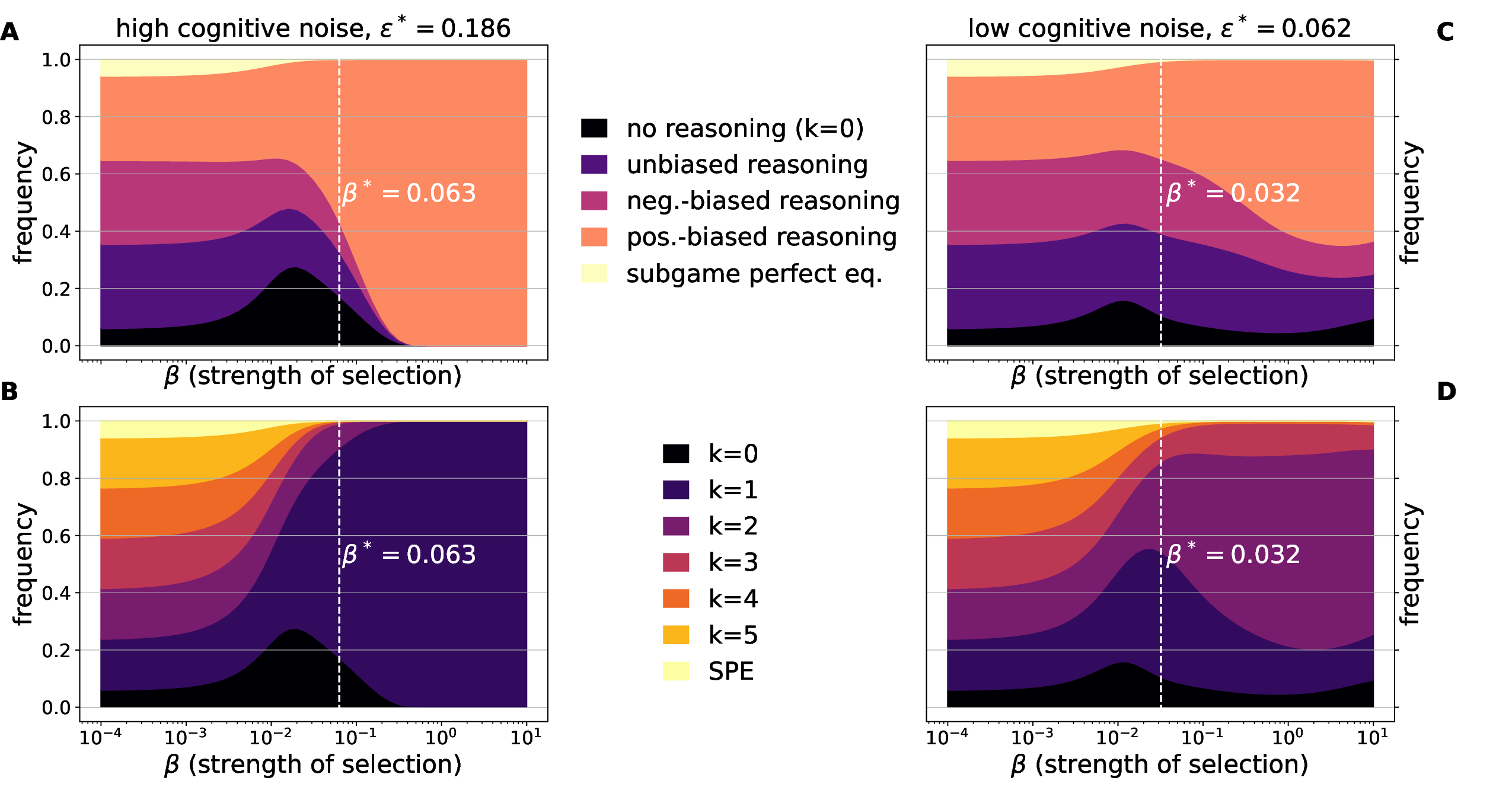}
}
\caption{Positively biased reasoning co-evolves with bounded rationality.
The distribution of reasoning types and reasoning levels in the six-step ICG is shown for variable selection strength $\beta$, under high ($\varepsilon^*=0.186$) and low ($\varepsilon^*=0.062$) probability of reasoning errors (panels \textbf{A,B} and \textbf{C,D}, respectively). 
For each value of $\beta$, we compute the stationary distribution $\phi$ of our dynamical system under the assumption of rare mutations. 
The frequencies of each strategy are aggregated by reasoning kernel, $f_{ker}=\sum_{k=1}^{L}\phi_{s_{ker, k}}$ for $ker\in \left\{U, B^{+}, B^{-}\right\}$, and by reasoning level, $f_{k}=\sum_{ker\in \left\{U, B^{+}, B^{-}\right\}}\phi_{s_{ker, k}}$ for $1\le k \le L$ (panels \textbf{A,C} and \textbf{B,D}, respectively).
For completeness, the figures also include the frequency of the pure strategy associated with no reasoning (i.e., reasoning level $k=0$) and the sub-game perfect equilibrium (SPE) strategy of stopping the game as early as possible.
The white dotted lines correspond to the selection strength $\beta^*$ of optimal fitting with the data in~\cite{kawagoe_level-k_2012} and~\cite{mckelvey_experimental_1992} (panels \textbf{A,B} and \textbf{C,D}, respectively). 
The population size is fixed to 100 individuals.
}
\label{fig:effect-of-beta}
\end{figure}

\subsection{Tables}
\begin{table}[!ht]
\centering
\resizebox{\textwidth}{!}{%
\begin{tabular}{c|c|c|c|c|c|c}
     & $k=0$                             & $k=1$        & $k=2$        & $k=3$        & $k=4$        & $k=5$        \\ \hline
Pl.1 & $\argmax_{0\le t\le 6}\pi_1(t)=6$ & $\BR_1(5)=4$ & $\BR_1(5)=4$ & $\BR_1(3)=2$ & $\BR_1(3)=2$ & $\BR_1(1)=0$ \\
Pl.2 & $\argmax_{0\le t\le 6}\pi_2(t)=5$ & $\BR_2(6)=5$ & $\BR_2(4)=3$ & $\BR_2(4)=3$ & $\BR_2(2)=1$ & $\BR_2(2)=1$
\end{tabular}
}
\caption{Representation of backward induction in a six-step Centipede Game as a level--$k$ recursive reasoning process.
From left to right, each column contains the actions of Player 1 (top) and Player 2 (bottom) as they are determined by a deterministic introspective process with an increasing number of reasoning steps. 
Such process starts at $k=0$ (left), where players do not perform any reasoning and will simply end the game in correspondence of the highest personal payoff.
For $k>0$, players choose the best response to a fictional level--$(k-1)$ player.
When $k=5$ (right), players perform the full backward induction process, thus the action generated by the reasoning process corresponds to the outcome predicted by the subgame perfect Nash equilibrium (SPE).}
\label{tab:backwardind}
\end{table}


\clearpage

\bibliographystyle{RS} 
\bibliography{references} 

\begin{thebibliography}{99}

\bibitem{camerer_behavioral_2011}
Camerer CF. 2011 {\em Behavioral Game Theory: Experiments in Strategic
  Interaction}.
Roundtable Series in Behavioral Economics. Princeton, NJ: Princeton University
  Press.
Reprint edition.

\bibitem{mckelvey_experimental_1992}
McKelvey RD, Palfrey TR. 1992  An {Experimental} {Study} of the {Centipede}
  {Game}. {\em Econometrica} \textbf{60}, 803.
(\href{http://dx.doi.org/10.2307/2951567}{10.2307/2951567})

\bibitem{rand_social_2014}
Rand DG, Peysakhovich A, Kraft-Todd GT, Newman GE, Wurzbacher O, Nowak MA,
  Greene JD. 2014  Social heuristics shape intuitive cooperation. {\em Nature
  Communications} \textbf{5}, 3677.
(\href{http://dx.doi.org/10.1038/ncomms4677}{10.1038/ncomms4677})

\bibitem{oosterbeek_cultural_2004}
Oosterbeek H, Sloof R, Van De~Kuilen G. 2004  Cultural {Differences} in
  {Ultimatum} {Game} {Experiments}: {Evidence} from a {Meta}-{Analysis}. {\em
  Experimental Economics} \textbf{7}, 171--188.
(\href{http://dx.doi.org/10.1023/B:EXEC.0000026978.14316.74}{10.1023/B:EXEC.0000026978.14316.74})

\bibitem{grosskopf_two-person_2008}
Grosskopf B, Nagel R. 2008  The two-person beauty contest. {\em Games and
  Economic Behavior} \textbf{62}, 93--99.
(\href{http://dx.doi.org/10.1016/j.geb.2007.03.004}{10.1016/j.geb.2007.03.004})

\bibitem{camerer2004behavioural}
Camerer CF, Ho TH, Chong JK. 2004 pp. 120--180.
In {\em Behavioural Game Theory: Thinking, Learning and Teaching}, pp.
  120--180. London: Palgrave Macmillan UK.
(\href{http://dx.doi.org/10.1057/9780230523371\_8}{10.1057/9780230523371\_8})

\bibitem{jensen_chimpanzees_2007}
Jensen K, Call J, Tomasello M. 2007  Chimpanzees {Are} {Rational} {Maximizers}
  in an {Ultimatum} {Game}. {\em Science} \textbf{318}, 107--109.
(\href{http://dx.doi.org/10.1126/science.1145850}{10.1126/science.1145850})

\bibitem{bornstein_rationality_2001}
Bornstein BH, Emler AC. 2001  Rationality in medical decision making: a review
  of the literature on doctors’ decision‐making biases. {\em Journal of
  Evaluation in Clinical Practice} \textbf{7}, 97--107.
(\href{http://dx.doi.org/10.1046/j.1365-2753.2001.00284.x}{10.1046/j.1365-2753.2001.00284.x})

\bibitem{montibeller_cognitive_2015}
Montibeller G, Von~Winterfeldt D. 2015  Cognitive and {Motivational} {Biases}
  in {Decision} and {Risk} {Analysis}. {\em Risk Analysis} \textbf{35},
  1230--1251.
(\href{http://dx.doi.org/10.1111/risa.12360}{10.1111/risa.12360})

\bibitem{marshall_evolutionary_2013}
Marshall JA, Trimmer PC, Houston AI, McNamara JM. 2013  On evolutionary
  explanations of cognitive biases. {\em Trends in Ecology \& Evolution}
  \textbf{28}, 469--473.
(\href{http://dx.doi.org/10.1016/j.tree.2013.05.013}{10.1016/j.tree.2013.05.013})

\bibitem{trivers_deceit_2011}
Trivers R. 2011 {\em Deceit and self-deception: fooling yourself the better to
  fool others}.
London: Allen Lane.

\bibitem{von_hippel_evolution_2011}
Von~Hippel W, Trivers R. 2011  The evolution and psychology of self-deception.
  {\em Behavioral and Brain Sciences} \textbf{34}, 1--16.
(\href{http://dx.doi.org/10.1017/S0140525X10001354}{10.1017/S0140525X10001354})

\bibitem{rusch_theory_2020}
Rusch T, Steixner-Kumar S, Doshi P, Spezio M, Gläscher J. 2020  Theory of mind
  and decision science: {Towards} a typology of tasks and computational models.
  {\em Neuropsychologia} \textbf{146}, 107488.
(\href{http://dx.doi.org/10.1016/j.neuropsychologia.2020.107488}{10.1016/j.neuropsychologia.2020.107488})

\bibitem{mckay_evolution_2009}
McKay RT, Dennett DC. 2009  The evolution of misbelief. {\em Behavioral and
  Brain Sciences} \textbf{32}, 493--510.
(\href{http://dx.doi.org/10.1017/S0140525X09990975}{10.1017/S0140525X09990975})

\bibitem{lenaerts_evolution_2024}
Lenaerts T, Saponara M, Pacheco JM, Santos FC. 2024  Evolution of a theory of
  mind. {\em iScience} \textbf{27}, 108862.
(\href{http://dx.doi.org/10.1016/j.isci.2024.108862}{10.1016/j.isci.2024.108862})

\bibitem{camerer_cognitive_2004}
Camerer CF, Ho TH, Chong JK. 2004  A {Cognitive} {Hierarchy} {Model} of
  {Games}. {\em The Quarterly Journal of Economics} \textbf{119}, 861--898.
(\href{http://dx.doi.org/10.1162/0033553041502225}{10.1162/0033553041502225})

\bibitem{smith_logic_1973}
Smith JM, Price GR. 1973  The {Logic} of {Animal} {Conflict}. {\em Nature}
  \textbf{246}, 15--18.
(\href{http://dx.doi.org/10.1038/246015a0}{10.1038/246015a0})

\bibitem{smith_evolution_1982}
Smith JM. 1982 {\em Evolution and the Theory of Games}.
Cambridge University Press.

\bibitem{sigmund_calculus_2010}
Sigmund K. 2010 {\em The calculus of selfishness}.
Princeton series in theoretical and computational biology. Princeton: Princeton
  University Press.
OCLC: ocn319157195.

\bibitem{fernandez_domingos_egttools_2023}
Fernández~Domingos E, Santos FC, Lenaerts T. 2023  {EGTtools}: {Evolutionary}
  game dynamics in {Python}. {\em iScience} \textbf{26}, 106419.
(\href{http://dx.doi.org/10.1016/j.isci.2023.106419}{10.1016/j.isci.2023.106419})

\bibitem{stahl_evolution_1993}
Stahl DO. 1993  Evolution of {Smartn} {Players}. {\em Games and Economic
  Behavior} \textbf{5}, 604--617.
(\href{http://dx.doi.org/10.1006/game.1993.1033}{10.1006/game.1993.1033})

\bibitem{kawagoe_level-k_2012}
Kawagoe T, Takizawa H. 2012  Level-k analysis of experimental centipede games.
  {\em Journal of Economic Behavior \& Organization} \textbf{82}, 548--566.
(\href{http://dx.doi.org/10.1016/j.jebo.2012.03.010}{10.1016/j.jebo.2012.03.010})

\bibitem{nax_deep_2022}
Nax HH, Newton J. 2022  Deep and shallow thinking in the long run. {\em
  Theoretical Economics} \textbf{17}, 1501--1527.
(\href{http://dx.doi.org/10.3982/TE4824}{10.3982/TE4824})

\bibitem{aumann_rationality_1997}
Aumann RJ. 1997  Rationality and {Bounded} {Rationality}. {\em Games and
  Economic Behavior} \textbf{21}, 2--14.
(\href{http://dx.doi.org/10.1006/game.1997.0585}{10.1006/game.1997.0585})

\bibitem{rosenthal_games_1981}
Rosenthal RW. 1981  Games of perfect information, predatory pricing and the
  chain-store paradox. {\em Journal of Economic Theory} \textbf{25}, 92--100.
(\href{http://dx.doi.org/10.1016/0022-0531(81)90018-1}{10.1016/0022-0531(81)90018-1})

\bibitem{rapoport_equilibrium_2003}
Rapoport A, Stein WE, Parco JE, Nicholas TE. 2003  Equilibrium play and
  adaptive learning in a three-person centipede game. {\em Games and Economic
  Behavior} \textbf{43}, 239--265.
(\href{http://dx.doi.org/10.1016/S0899-8256(03)00009-5}{10.1016/S0899-8256(03)00009-5})

\bibitem{murphy_population_2004}
Murphy RO, Rapoport A, Parco JE. 2004  Population {Learning} of {Cooperative}
  {Behavior} in a {Three}-{Person} {Centipede} {Game}. {\em Rationality and
  Society} \textbf{16}, 91--120.
(\href{http://dx.doi.org/10.1177/1043463104039876}{10.1177/1043463104039876})

\bibitem{fey_experimental_1996}
Fey M, McKelvey RD, Palfrey TR. 1996  An experimental study of constant-sum
  centipede games. {\em International Journal of Game Theory} \textbf{25},
  269--287.
(\href{http://dx.doi.org/10.1007/BF02425258}{10.1007/BF02425258})

\bibitem{palacios-huerta_field_2009}
Palacios-Huerta I, Volij O. 2009  Field {Centipedes}. {\em American Economic
  Review} \textbf{99}, 1619--1635.
(\href{http://dx.doi.org/10.1257/aer.99.4.1619}{10.1257/aer.99.4.1619})

\bibitem{rand_evolutionary_2012}
Rand DG, Nowak MA. 2012  Evolutionary dynamics in finite populations can
  explain the full range of cooperative behaviors observed in the centipede
  game. {\em Journal of Theoretical Biology} \textbf{300}, 212--221.
(\href{http://dx.doi.org/10.1016/j.jtbi.2012.01.011}{10.1016/j.jtbi.2012.01.011})

\bibitem{mckelvey_quantal_1995}
McKelvey RD, Palfrey TR. 1995  Quantal {Response} {Equilibria} for {Normal}
  {Form} {Games}. {\em Games and Economic Behavior} \textbf{10}, 6--38.
(\href{http://dx.doi.org/10.1006/game.1995.1023}{10.1006/game.1995.1023})

\bibitem{mckelvey_quantal_1998}
Mckelvey RD, Palfrey TR. 1998  Quantal {Response} {Equilibria} for {Extensive}
  {Form} {Games}. {\em Experimental Economics} \textbf{1}, 9--41.
(\href{http://dx.doi.org/10.1023/A:1009905800005}{10.1023/A:1009905800005})

\bibitem{bela_altruism_2022}
Béla E. 2022  Altruism and {Ambiguity} in the {Centipede} game.
  (\href{http://dx.doi.org/10.31235/osf.io/93p8s}{10.31235/osf.io/93p8s})

\bibitem{krockow_exploring_2016}
Krockow EM, Colman AM, Pulford BD. 2016  Exploring cooperation and competition
  in the {Centipede} game through verbal protocol analysis. {\em European
  Journal of Social Psychology} \textbf{46}, 746--761.
(\href{http://dx.doi.org/10.1002/ejsp.2226}{10.1002/ejsp.2226})

\bibitem{gamba_preferences-dependent_2015}
Gamba A, Regner T. 2019  Preferences-dependent learning in the centipede game:
  The persistence of mistrust. {\em European Economic Review} \textbf{120},
  103316.
(\href{http://dx.doi.org/https://doi.org/10.1016/j.euroecorev.2019.103316}{https://doi.org/10.1016/j.euroecorev.2019.103316})

\bibitem{fawcett_evolution_2014}
Fawcett TW, Fallenstein B, Higginson AD, Houston AI, Mallpress DE, Trimmer PC,
  McNamara JM. 2014  The evolution of decision rules in complex environments.
  {\em Trends in Cognitive Sciences} \textbf{18}, 153--161.
(\href{http://dx.doi.org/10.1016/j.tics.2013.12.012}{10.1016/j.tics.2013.12.012})

\bibitem{tversky_judgment_1974}
Tversky A, Kahneman D. 1974  Judgment under {Uncertainty}: {Heuristics} and
  {Biases}: {Biases} in judgments reveal some heuristics of thinking under
  uncertainty.. {\em Science} \textbf{185}, 1124--1131.
(\href{http://dx.doi.org/10.1126/science.185.4157.1124}{10.1126/science.185.4157.1124})

\bibitem{de_weerd_how_2013}
De~Weerd H, Verbrugge R, Verheij B. 2013  How much does it help to know what
  she knows you know? {An} agent-based simulation study. {\em Artificial
  Intelligence} \textbf{199-200}, 67--92.
(\href{http://dx.doi.org/10.1016/j.artint.2013.05.004}{10.1016/j.artint.2013.05.004})

\bibitem{devaine_theory_2014}
Devaine M, Hollard G, Daunizeau J. 2014  Theory of {Mind}: {Did} {Evolution}
  {Fool} {Us}?. {\em PLoS ONE} \textbf{9}, e87619.
(\href{http://dx.doi.org/10.1371/journal.pone.0087619}{10.1371/journal.pone.0087619})

\bibitem{basu_travelers_1994}
Basu K. 1994  The {Traveler}'s {Dilemma}: {Paradoxes} of {Rationality} in
  {Game} {Theory}. {\em The American Economic Review} \textbf{84}, 391--395.

\bibitem{goeree_stochastic_1999}
Goeree JK, Holt CA. 1999  Stochastic game theory: {For} playing games, not just
  for doing theory. {\em Proceedings of the National Academy of Sciences}
  \textbf{96}, 10564--10567.
(\href{http://dx.doi.org/10.1073/pnas.96.19.10564}{10.1073/pnas.96.19.10564})

\bibitem{goeree_model_2004}
Goeree JK, Holt CA. 2004  A model of noisy introspection. {\em Games and
  Economic Behavior} \textbf{46}, 365--382.
(\href{http://dx.doi.org/10.1016/S0899-8256(03)00145-3}{10.1016/S0899-8256(03)00145-3})

\bibitem{stewart_extortion_2013}
Stewart AJ, Plotkin JB. 2013  From extortion to generosity, evolution in the
  {Iterated} {Prisoner}’s {Dilemma}. {\em Proceedings of the National Academy
  of Sciences} \textbf{110}, 15348--15353.
(\href{http://dx.doi.org/10.1073/pnas.1306246110}{10.1073/pnas.1306246110})

\bibitem{auger_hawk-dove_1998}
Auger P, De~La~Parra R, Sánchez E. 1998  Hawk-dove game and competition
  dynamics. {\em Mathematical and Computer Modelling} \textbf{27}, 89--98.
(\href{http://dx.doi.org/10.1016/S0895-7177(98)00009-0}{10.1016/S0895-7177(98)00009-0})

\bibitem{schuster1983replicator}
Schuster P, Sigmund K. 1983  Replicator dynamics. {\em Journal of theoretical
  biology} \textbf{100}, 533--538.

\bibitem{cressman2014replicator}
Cressman R, Tao Y. 2014  The replicator equation and other game dynamics. {\em
  Proceedings of the National Academy of Sciences} \textbf{111}, 10810--10817.

\bibitem{nowak2006five}
Nowak MA. 2006  Five rules for the evolution of cooperation. {\em science}
  \textbf{314}, 1560--1563.

\bibitem{taylor2007transforming}
Taylor C, Nowak MA. 2007  Transforming the dilemma. {\em Evolution}
  \textbf{61}, 2281--2292.

\bibitem{heller_three_2015}
Heller Y. 2015  Three steps ahead: {Three} steps ahead. {\em Theoretical
  Economics} \textbf{10}, 203--241.
(\href{http://dx.doi.org/10.3982/TE1660}{10.3982/TE1660})

\bibitem{crawford_lying_2003}
Crawford VP. 2003  Lying for {Strategic} {Advantage}: {Rational} and
  {Boundedly} {Rational} {Misrepresentation} of {Intentions}. {\em American
  Economic Review} \textbf{93}, 133--149.
(\href{http://dx.doi.org/10.1257/000282803321455197}{10.1257/000282803321455197})

\bibitem{aue_neural_2012}
Aue T, Nusbaum HC, Cacioppo JT. 2012  Neural correlates of wishful thinking.
  {\em Social Cognitive and Affective Neuroscience} \textbf{7}, 991--1000.
(\href{http://dx.doi.org/10.1093/scan/nsr081}{10.1093/scan/nsr081})

\bibitem{neumann_bounded_2014}
Neuman R, Rafferty A, Griffiths T. 2014  A bounded rationality account of
  wishful thinking. In {\em Proceedings of the annual meeting of the Cognitive
  Science Society} vol.~36.

\bibitem{yildiz_wishful_2007}
Yildiz M. 2007  Wishful Thinking in Strategic Environments. {\em The Review of
  Economic Studies} \textbf{74}, 319--344.
(\href{http://dx.doi.org/10.1111/j.1467-937X.2007.00423.x}{10.1111/j.1467-937X.2007.00423.x})

\bibitem{vosgerau_how_2010}
Vosgerau J. 2010  How prevalent is wishful thinking? {Misattribution} of
  arousal causes optimism and pessimism in subjective probabilities.. {\em
  Journal of Experimental Psychology: General} \textbf{139}, 32--48.
(\href{http://dx.doi.org/10.1037/a0018144}{10.1037/a0018144})

\bibitem{sanna_defensive_1998}
Sanna LJ. 1998  Defensive {Pessimism} and {Optimism}: {The} {Bitter}-{Sweet}
  {Influence} of {Mood} on {Performance} and {Prefactual} and {Counterfactual}
  {Thinking}. {\em Cognition \& Emotion} \textbf{12}, 635--665.
(\href{http://dx.doi.org/10.1080/026999398379484}{10.1080/026999398379484})

\bibitem{bateson_agitated_2011}
Bateson M, Desire S, Gartside S, Wright G. 2011  Agitated {Honeybees} {Exhibit}
  {Pessimistic} {Cognitive} {Biases}. {\em Current Biology} \textbf{21},
  1070--1073.
(\href{http://dx.doi.org/10.1016/j.cub.2011.05.017}{10.1016/j.cub.2011.05.017})

\bibitem{krizan_wishful_2009}
Krizan Z, Windschitl PD. 2009  Wishful {Thinking} about the {Future}: {Does}
  {Desire} {Impact} {Optimism}?. {\em Social and Personality Psychology
  Compass} \textbf{3}, 227--243.
(\href{http://dx.doi.org/10.1111/j.1751-9004.2009.00169.x}{10.1111/j.1751-9004.2009.00169.x})

\bibitem{krizan_influence_2007}
Krizan Z, Windschitl PD. 2007  The influence of outcome desirability on
  optimism.. {\em Psychological Bulletin} \textbf{133}, 95--121.
(\href{http://dx.doi.org/10.1037/0033-2909.133.1.95}{10.1037/0033-2909.133.1.95})

\bibitem{dekel1999evolution}
Dekel E, Scotchmer S. 1999  On the Evolution of Attitudes towards Risk in
  Winner-Take-All Games. {\em Journal of Economic Theory} \textbf{87},
  125--143.
(\href{http://dx.doi.org/https://doi.org/10.1006/jeth.1999.2537}{https://doi.org/10.1006/jeth.1999.2537})

\bibitem{brocas2025children}
Brocas I, Carrillo JD. 2025  Why do children pass in the centipede game?
  Cognitive limitations v. risk calculations. {\em Games and Economic Behavior}
  \textbf{150}, 295--311.
(\href{http://dx.doi.org/https://doi.org/10.1016/j.geb.2025.01.003}{https://doi.org/10.1016/j.geb.2025.01.003})

\bibitem{johnson_evolution_2011}
Johnson DDP, Fowler JH. 2011  The evolution of overconfidence. {\em Nature}
  \textbf{477}, 317--320.
(\href{http://dx.doi.org/10.1038/nature10384}{10.1038/nature10384})

\bibitem{fatima_learning_2024}
Fatima S, Jennings NR, Wooldridge M. 2024  Learning to {Resolve} {Social}
  {Dilemmas}: {A} {Survey}. {\em Journal of Artificial Intelligence Research}
  \textbf{79}, 895--969.
(\href{http://dx.doi.org/10.1613/jair.1.15167}{10.1613/jair.1.15167})

\bibitem{fudenberg_imitation_2006}
Fudenberg D, Imhof LA. 2006  Imitation processes with small mutations. {\em
  Journal of Economic Theory} \textbf{131}, 251--262.
(\href{http://dx.doi.org/10.1016/j.jet.2005.04.006}{10.1016/j.jet.2005.04.006})

\bibitem{traulsen_stochastic_2006}
Traulsen A, Nowak MA, Pacheco JM. 2006  Stochastic dynamics of invasion and
  fixation. {\em Physical Review E} \textbf{74}, 011909.
(\href{http://dx.doi.org/10.1103/PhysRevE.74.011909}{10.1103/PhysRevE.74.011909})

\bibitem{hindersin_computation_2019}
Hindersin L, Wu B, Traulsen A, García J. 2019  Computation and {Simulation} of
  {Evolutionary} {Game} {Dynamics} in {Finite} {Populations}. {\em Scientific
  Reports} \textbf{9}, 6946.
(\href{http://dx.doi.org/10.1038/s41598-019-43102-z}{10.1038/s41598-019-43102-z})

\bibitem{marco_saponara.2025.15838520}
Saponara M. 2025  MarcoSaponara/centipede-bias: First public release.
  (\href{http://dx.doi.org/10.5281/zenodo.15838520}{10.5281/zenodo.15838520})

\bibitem{yamagishi_trust_1994}
Yamagishi T, Yamagishi M. 1994  Trust and commitment in the {United} {States}
  and {Japan}. {\em Motivation and Emotion} \textbf{18}, 129--166.
(\href{http://dx.doi.org/10.1007/BF02249397}{10.1007/BF02249397})

\bibitem{krockow_commitment-enhancing_2018}
Krockow EM, Takezawa M, Pulford BD, Colman AM, Smithers S, Kita T, Nakawake Y.
  2018  Commitment-enhancing tools in {Centipede} games: {Evidencing}
  {European}–{Japanese} differences in trust and cooperation. {\em Judgment
  and Decision Making} \textbf{13}, 61--72.
(\href{http://dx.doi.org/10.1017/S1930297500008822}{10.1017/S1930297500008822})

\bibitem{krockow_cooperation_2016}
Krockow EM, Colman AM, Pulford BD. 2016  Cooperation in repeated interactions:
  {A} systematic review of {Centipede} game experiments, 1992–2016. {\em
  European Review of Social Psychology} \textbf{27}, 231--282.
(\href{http://dx.doi.org/10.1080/10463283.2016.1249640}{10.1080/10463283.2016.1249640})

\bibitem{krockow_cooperation_2017}
Krockow EM, Takezawa M, Pulford BD, Colman AM, Kita T. 2017  Cooperation and
  {Trust} in {Japanese} and {British} {Samples}: {Evidence} {From} {Incomplete}
  {Information} {Games}. {\em International Perspectives in Psychology}
  \textbf{6}, 227--245.
(\href{http://dx.doi.org/10.1037/ipp0000074}{10.1037/ipp0000074})

\end{thebibliography}

\appendix

\section{Fitting with experimental data}\label{sec:appendixA}
We address the choice of the pair of values for the parameters of our model, namely the strength of selection $\beta$ and the probability of reasoning error $\varepsilon$.
In order to retrieve reasonable values, we fitted our model to two different experimental references already mentioned in the Results: the seminal study by McKelvey and Palfrey~\cite{mckelvey_experimental_1992} and the more recent work by Kawagoe and Takizawa~\cite{kawagoe_level-k_2012}. 
Both references provide a benchmark for the behavioural profile in the ICG with $L=6$ depicted in \autoref{fig:centipede}. 
The main difference between these studies relies on the fact that the latter behavioural experiment is a one-shot interaction, while the former consists of ten iterations with random pairing (in this case, we exclusively refer to the frequencies associated with the first round, to eliminate potential learning effects). 
Cultural differences may also be at play~\cite{yamagishi_trust_1994, krockow_commitment-enhancing_2018, krockow_cooperation_2016, krockow_cooperation_2017}, as the experiments in Refs.~\cite{mckelvey_experimental_1992} and~\cite{kawagoe_level-k_2012} were conducted in American and Japanese universities, respectively.

\begin{figure}
\centering
\resizebox{\textwidth}{!}{%
\includegraphics{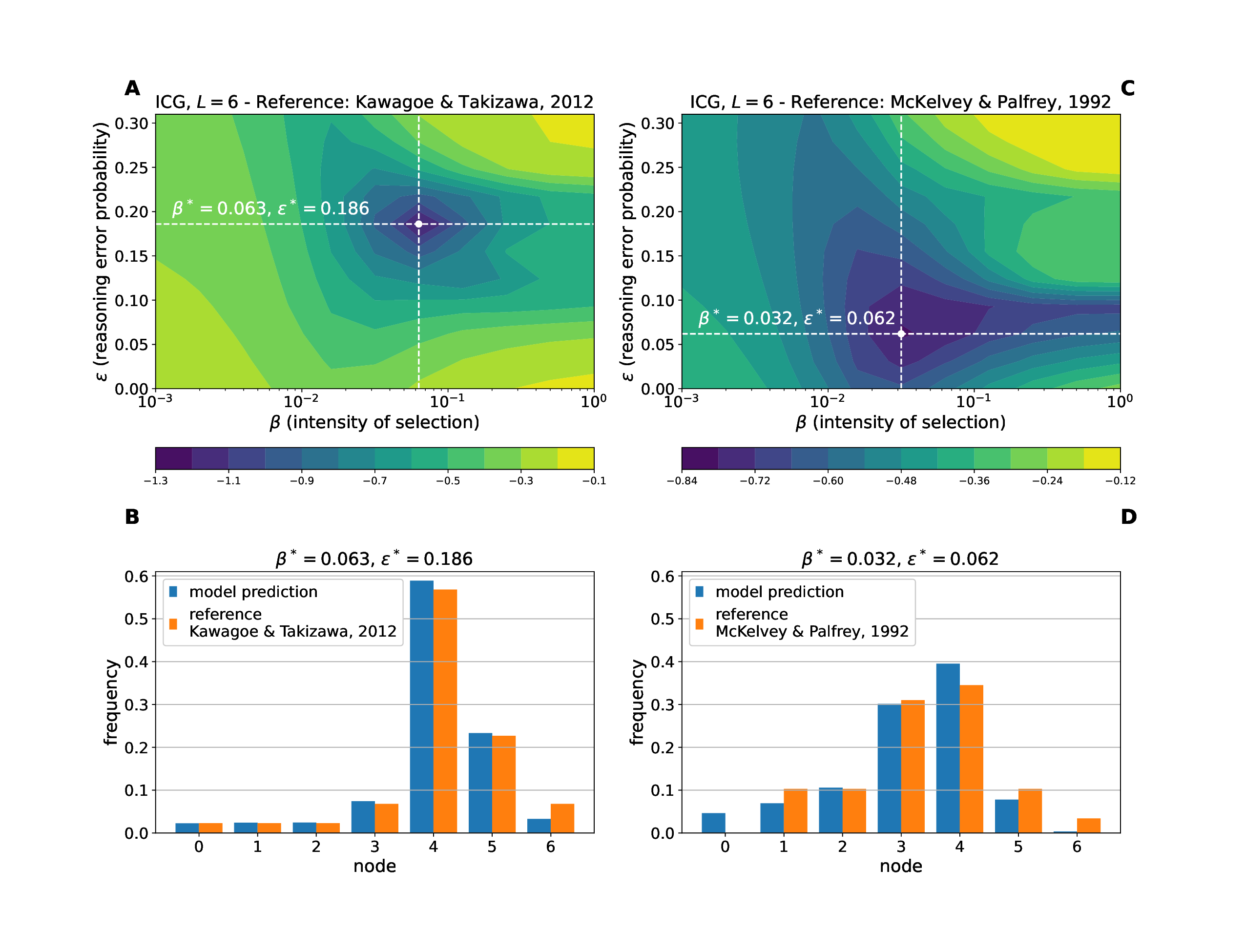}
}
\caption{Fitting with experimental references. 
We fit our model to the experimental data related to the ICG with $L=6$ from Refs.~\cite{kawagoe_level-k_2012} and~\cite{mckelvey_experimental_1992} (panels \textbf{A,B} and \textbf{C,D}, respectively).
Panels \textbf{A,C} show the Jensen-Shannon divergence between our model's prediction and each experimental reference (in logarithmic scale) as a function of the strength of selection $\beta$ and the probability of reasoning errors $\varepsilon$.
In the case of Ref.~\cite{kawagoe_level-k_2012}, the minimum distance (0.06) is achieved when $\beta^*=0.063$ and $\varepsilon^*=0.186$.
When using the data from Ref.~\cite{mckelvey_experimental_1992}, the minimum distance (0.16) is reached when $\beta^*=0.032$ and $\varepsilon^*=0.062$.
Panels \textbf{B,D} portray the comparison between the optimal fitting and the experimental reference in both cases.
The population size is fixed to 100 individuals.
}
\label{fig:fitting}
\end{figure}

These differences are reflected in the results of the fitting shown in \autoref{fig:fitting}.
By minimizing the Jensen-Shannon divergence between our model's prediction of the terminal node of the game and the two experimental references, we obtain rather different results.
When using the data from Ref.~\cite{kawagoe_level-k_2012} (Panel \textbf{A}), the best fit is located where the strength of selection and the cognitive noise take values $\beta^*=0.063$ and $\varepsilon^*=0.186$, respectively. 
In this case, the relatively high value of $\varepsilon$ implies that the three different reasoning kernels, and particularly the positively biased one, are necessary to explain the experimental results.

In contrast, when using the data from Ref.~\cite{mckelvey_experimental_1992} (Panel \textbf{B}), the best fitting is given by lower values, $\beta^*=0.032$ and $\varepsilon^*=0.062$ respectively. 
This implies that, in this case, the most relevant factors to explain the data are the stochastic effects of the imitation process (represented by the low strength of selection), as well as the reasoning capacity of each individual (represented by the level $k$).
However, the low value of cognitive noise leads to marginal differences in reasoning types and consequently to almost deterministic strategies.
Despite the highly mitigated differences between reasoning kernels, we have seen in \autoref{fig:effect-of-beta} (Panel \textbf{C}) that it is yet preferable to develop positively biased reasoning, especially under the effect of stronger selection.

Although \textit{general trust}, intended as an expectation of strangers' goodwill, has been suggested to be more prevalent in western cultures (particularly Anglophone), Japanese culture has been associated with \textit{assurance-based trust}, which does not require a belief in the benevolence of the other person because it is rather based on the mutual knowledge of the incentive structure surrounding the relationship~\cite{yamagishi_trust_1994}. 
This general finding is confirmed by cross-cultural studies involving the Centipede Game~\cite{krockow_cooperation_2016, krockow_cooperation_2017}, where Japanese participants systematically ended the game at nodes later than Western participants.
Our results appear to corroborate these previous findings, as approximately 51\% of the population in our model adopts the strategy associated with one step of positively biased reasoning when fitted on the Japanese data, while the distribution of strategies is more heterogeneous when the model is fitted on the American data, with one step of positively biased reasoning being nevertheless the most frequent strategy (approximately 18\%). 

\section{Derivation of the reasoning strategies}\label{sec:appendixB}
Here we report the mathematical derivation of the strategies of our model.
Starting from \autoref{eq:totalprob}, it is useful to encode the noisy best responses, $\NBR_i(\cdot)$, into two role-dependent matrices, $M_1(\varepsilon)$ and $M_2(\varepsilon)$, which depend on the intensity of the cognitive noise $\varepsilon$ and contain the probabilities of choosing a certain response (column) given the belief about the action of the co-player (row):
\begin{eqnarray}
    \left[M_i(\varepsilon)\right]_{t_{-i}, t_{i}} = 
    \mathbb{P}(\NBR_i(t_{-i})=t_i\ |\ T_{-i}(k-1)=t_{-i}).\notag
\end{eqnarray}
Then, we can rewrite \autoref{eq:totalprob} in the following compact form:
\begin{eqnarray}
    \sigma_i(k) = \sigma_{-i}(k-1)\cdot M_i(\varepsilon)\notag
\end{eqnarray}
with $i\in\left\{1,2\right\}$, and $k>0$. 
Finally, if we concatenate the two mixed strategy vectors $\sigma_1(k)$ and $\sigma_2(k)$ into one vector $\sigma(k) = \left[\sigma_{1}(k),\ \sigma_{2}(k)\right]$ and define a single $(L+2)\times(L+2)$ block matrix, $M(\varepsilon)$, such that
\begin{eqnarray}
    M(\varepsilon) = \begin{bmatrix}
        \textbf{0} & M_2(\varepsilon)\\
        M_1(\varepsilon) & \textbf{0}
    \end{bmatrix}\notag
\end{eqnarray}
we obtain the following simple analytical formula encompassing the strategy of a player covering both roles with equal probability:
\begin{eqnarray}
    \sigma(k) = \sigma(0)\cdot M(\varepsilon)^{k}.\notag
\end{eqnarray}

As mentioned in the Methods, we consider three reasoning kernels, namely $M_{U}(\varepsilon)$, $M_{B^{-}}(\varepsilon)$, and $M_{B^{+}}(\varepsilon)$ for unbiased, negatively and positively biased reasoning, respectively. 
Here, we report the three matrices for the case $L=6$:
\begin{align}
    M_{U}(\varepsilon) = \begin{bmatrix}
        0 & 0 & 0 & 0 & 1/4 & 1/4 & 1/4 & 1/4\\
        0 & 0 & 0 & 0 & 1-\varepsilon & \varepsilon/3 & \varepsilon/3 & \varepsilon/3\\
        0 & 0 & 0 & 0 & \varepsilon/3 & 1-\varepsilon & \varepsilon/3 & \varepsilon/3\\
        0 & 0 & 0 & 0 & \varepsilon/3 & \varepsilon/3 & 1-\varepsilon & \varepsilon/3\\
        1-\varepsilon & \varepsilon/3 & \varepsilon/3 & \varepsilon/3 & 0 & 0 & 0 & 0\\
        \varepsilon/3 & 1-\varepsilon & \varepsilon/3 & \varepsilon/3 & 0 & 0 & 0 & 0\\
        \varepsilon/3 & \varepsilon/3 & 1-\varepsilon & \varepsilon/3 & 0 & 0 & 0 & 0\\
        \varepsilon/3 & \varepsilon/3 & \varepsilon/3 & 1-\varepsilon & 0 & 0 & 0 & 0
    \end{bmatrix},\notag\\
    M_{B^{-}}(\varepsilon) = \begin{bmatrix}
        0 & 0 & 0 & 0 & 1/4 & 1/4 & 1/4 & 1/4\\
        0 & 0 & 0 & 0 & 1 & 0 & 0 & 0\\
        0 & 0 & 0 & 0 & \varepsilon & 1-\varepsilon & 0 & 0\\
        0 & 0 & 0 & 0 & \varepsilon/2 & \varepsilon/2 & 1-\varepsilon & 0\\
        1 & 0 & 0 & 0 & 0 & 0 & 0 & 0\\
        \varepsilon & 1-\varepsilon & 0 & 0 & 0 & 0 & 0 & 0\\
        \varepsilon/2 & \varepsilon/2 & 1-\varepsilon & 0 & 0 & 0 & 0 & 0\\
        \varepsilon/3 & \varepsilon/3 & \varepsilon/3 & 1-\varepsilon & 0 & 0 & 0 & 0
    \end{bmatrix},\notag\\
    M_{B^{+}}(\varepsilon) = \begin{bmatrix}
        0 & 0 & 0 & 0 & 1/4 & 1/4 & 1/4 & 1/4\\
        0 & 0 & 0 & 0 & 1-\varepsilon & \varepsilon/3 & \varepsilon/3 & \varepsilon/3\\
        0 & 0 & 0 & 0 & 0 & 1-\varepsilon & \varepsilon/2 & \varepsilon/2\\
        0 & 0 & 0 & 0 & 0 & 0 & 1-\varepsilon & \varepsilon\\
        1-\varepsilon & \varepsilon/3 & \varepsilon/3 & \varepsilon/3 & 0 & 0 & 0 & 0\\
        0 & 1-\varepsilon & \varepsilon/2 & \varepsilon/2 & 0 & 0 & 0 & 0\\
        0 & 0 & 1-\varepsilon & \varepsilon & 0 & 0 & 0 & 0\\
        0 & 0 & 0 & 1 & 0 & 0 & 0 & 0
    \end{bmatrix}.\notag
\end{align}
Note that if Player 2 believes that Player 1 will stop the game at the first node ($T=0$), then Player 2 is indifferent about the action to take, i.e., there is no best response. This explains the uniform probability distribution over the actions of Player 2 in the first row of each matrix.

\end{document}